\documentclass[preprint]{iucr}              

     \journalcode{S}              

\usepackage{amsmath}
\usepackage{graphicx}
\usepackage{datatool}
\usepackage{multicol}
\usepackage{hyperref}
\def\equationautorefname~#1\null{%
  Equation~(#1)\null
}
\usepackage{todonotes}
\usepackage{tabularx}
\usepackage{soul}
\begin{document}

\title{Evaluation of X-ray/EUV Nanolithography Facility at AS Through Wavefront Propagation Simulations}

\cauthor[a]{Jerome}{B.M. Knappett}{j.knappett@latrobe.edu.au}{}
\author[a]{Blair}{Haydon}{}
\author[b]{Bruce}{C.C. Cowie}{}
\author[b,a]{Cameron}{M. Kewish}{}
\author[a]{Grant}{A. van Riessen}{}

\aff[a]{Department of Mathematical and Physical Sciences, School of Computing, Engineering and Mathematical Sciences, La Trobe University, Bundoora, Victoria 3086 \country{Australia}}
\aff[b]{Australian Synchrotron, Australian Nuclear Science and Technology Organisation (ANSTO), Clayton, Victoria 3168 \country{Australia}}


\maketitle

\begin{synopsis}
The optical performance and suitability for EUV interference lithography of the soft X-ray beamline of the Australian Synchrotron is evaluated through comparisons of partially coherent simulation and experimental measurements.
\end{synopsis}

\begin{abstract}
Synchrotron light sources can provide the required spatial coherence, stability and control that is required to support the development of advanced lithography at the extreme ultraviolet and soft X-ray wavelengths that are relevant to current and future fabricating technologies. Here we present an evaluation of the optical performance of the soft X-ray (SXR) beamline of the  Australian Synchrotron (AS) and its suitability for developing interference lithography using radiation in the 91.8~eV (13.5~nm) to 300~eV (4.13~nm) range. A comprehensive physical optics model of the APPLE-II undulator source and SXR beamline was constructed to simulate the properties of the illumination at the proposed location of a photomask, as a function of photon energy, collimation, and monochromator parameters. The model is validated using a combination of experimental measurements of the photon intensity distribution of the undulator harmonics. We show that the undulator harmonics intensity ratio can be accurately measured using an imaging detector and controlled using beamline optics.  Finally, we evaluate photomask geometric constraints and achievable performance for the limiting case of fully spatially coherent illumination.
\end{abstract}


\section{Introduction}
\label{s:Introduction}

As of 2019, the semiconductor device manufacturing industry has adopted lithography technology utilising EUV radiation of wavelength $\lambda=13.5$~nm, corresponding to photon energy 91.8~eV, for high-volume manufacturing (HVM) \cite{fomenkov_euv_2019}. To keep up with the demands of device scaling predicted by Moore's Law, a future transition to 6.7~nm wavelength (185~eV) sources is anticipated. The availability of laser pulsed plasma light sources \cite{otsuka_6_2012}, and multilayer optics \cite{uzoma_multilayer_2021} at this wavelength make it a particularly promising candidate for future lithography.  A shift to shorter wavelengths brings with it significant challenges, including 
an increased significance of stochastic effects for higher resolution patterning \cite{de_stochastic_2017}, and a need for understanding the wavelength-dependent effects of mask defects \cite{goldberg_wavelength-specific_2010}. Interference lithography (IL) using synchrotron radiation has recently emerged as a powerful tool for understanding the challenges for future lithographic processes, including photoresist performance from EUV \cite{mojarad_beyond_2015} to SXR wavelengths (2.5~nm)~\cite{mojarad_fabrication_2021}. Synchrotron sources are ideal for the development of lithographic technology as they allow for the control of properties such as flux, coherence and polarisation, and can cover the entire energy range from EUV to SXR.
Partially coherent wavefront propagation simulation can provide critical insight into the design of optical systems and EUV/SXR-IL process development using existing synchrotron radiation sources.

For high-resolution patterning using IL, it is critical to achieve high contrast in the aerial image formed where beams diffracted from two or more gratings interfere.  This requires that light sources used in IL provide high intensity,  spatial coherence, and stability \cite{mojarad_interference_2015}. The necessary spatial coherence length to ensure high contrast aerial images is determined by the size of the desired exposure area \cite{solak_sub-50_2003}, which is, in turn, defined by the maximum separation between points in a set of two or more radially arranged gratings. 
Fully coherent illuminaton of the grating set typically requires a coherence length 3$\times$ the linear dimensions of the desired exposure area \cite{solak_four-wave_2002}.
Typical grating masks used for EUV-IL 
require a lateral coherence length at the mask plane of between 150~\textmu m to 1.2~mm \cite{ekinci_euv_2014,meng_analysis_2021}, although small area patterning using EUV-IL has been reported using a coherence length of just 43.2~\textmu m \cite{sahoo_development_2023}.

When evaluating the performance of a light source for interference lithography, the relevant quantity that determines throughput is the exposure time required to provide the radiation dose-on-wafer $D_w(\lambda)$ to transfer a pattern to a film of photosensitive resist with particular sensitivity at a particular wavelength $\lambda$. This, in turn, depends on the coherent intensity at the mask and mask efficiency. Practically, for studies of lithographic processes at high resolution, throughput is of lower priority than lithographic quality, and the exposure time determines the required mechanical stability between the mask and wafer.

Photoresist sensitivity is defined by the dose $D_0(\lambda)$ \cite{ekinci_evaluation_2013, mojarad_beyond_2015} required to remove 50\% of the resist thickness during development \cite{ekinci_evaluation_2013} for a particular wavelength.  
Resist sensitivity generally decreases for shorter wavelengths, \textit{i.e.}, $D_0 \propto 1/\lambda$.  Combined with the requirement for higher doses for patterning thinner resists and smaller feature sizes, the scaling of source power requirements with industry device scaling targets is a major challenge~\cite{van_high_2021,levinson_high_2022}.
For the present work, it is convenient to define the dose-on-mask, $D_m$, required to achieve a target value of $D_0$, taking into account  diffraction efficiency, $\eta(\lambda)$, of the grating mask and the transmission of the substrate, $T(\lambda)$, at the relevant wavelength:
\begin{equation}
    D_m = \frac{D_0}{\eta T N^3},
    \label{e:MasktoWaferDose}
\end{equation}
where $N$ is the number of gratings contained in the mask and the power of 3 is due to the definition of the efficiency of an IL grating, $\eta_{IL}=N\eta$ \cite{mojarad_interference_2015}, and 
periodicity in the aerial image intensity. The wavelength dependence has been suppressed for simplicity. While diffraction efficiencies of up to 28\% have been reported for low resolution IL gratings \cite{braig_stand_2011}, Wang \textit{et al.}~\citeyear{wang_high_2021} reported diffraction efficiencies for a bilayer of HSQ and spin-on-carbon of $\eta>6$\% for gratings with HP down to 12~nm  at EUV wavelengths.

Organic chemically amplified resists (CARs) have been used for performance comparisons of EUV and SXR, as they offer high-resolution patterning in both wavelength ranges~\cite{mojarad_patterning_2013}. Mojarad \textit{et al.}~\citeyear{mojarad_beyond_2015} reported 
CAR sensitivities of $D_0(\text{13.5~nm})=11.6$~mJ/cm$^2$, and $D_0(\text{6.5~nm})=33.2$~mJ/cm$^2$. The 2021 International Roadmap for Devices and Systems has predicted that $D_0 > 80$ mJ/cm\textsuperscript{2} will be required for 10~nm half-pitch patterning \cite{levinson_high_2022}.  Assuming a SiO\textsubscript{2} substrate of 40~nm thickness, and a 
4-grating mask with $\eta=6$\% efficiency, the corresponding minimum dose on mask for EUV-IL can be estimated to be $D_m\sim11$~mJ/cm\textsuperscript{2}.  

In this work, partially coherent wavefront propagation was used to simulate the complex wavefield at the mask plane of an EUV-IL instrument being developed at La Trobe University for one of two branches of the SXR beamline of the Australian Synchrotron (AS). Simulations were performed primarily at 90.44~eV and 184.76~eV photon energy and the relevant characteristics of the beam were compared to direct measurements of the intensity taken at the beamline with photon energy ranging from 90~eV to 300~eV. We demonstrate, by comparison to photoelectron spectroscopy measurements, that the ratio of undulator harmonic intensity can be measured using an imaging detector. We discuss how the higher harmonic intensity scales with decreasing photon energy and is affected by collimating slits also used to control spatial coherence.  The beamline's suitability for EUV-IL is evaluated with respect to the compromise between the grating area that can be coherently illuminated and the intensity of the illumination.

\section{Theory and Modelling}
\label{s:Theory-And-Modelling}
\subsection{Interference Lithography}
\label{s:Interference-Lithography}

Light incident on a grating diffracts at multiple angles, $\theta_m$, corresponding to orders, $m$, governed by the grating equation. For a monochromatic beam of wavelength, $\lambda$, incident at angle, $\theta_i$, on a grating with period, $p_G$, the angle of diffraction of the $m^{th}$ order beam is \cite{hecht_optics_1987}
\begin{equation}
    \sin \theta_m = \frac{m \lambda}{p_G} + \sin \theta_i .
    \label{e:GratingEquation}
\end{equation}

\begin{figure}
    \centering
    \includegraphics[width=0.50\textwidth]{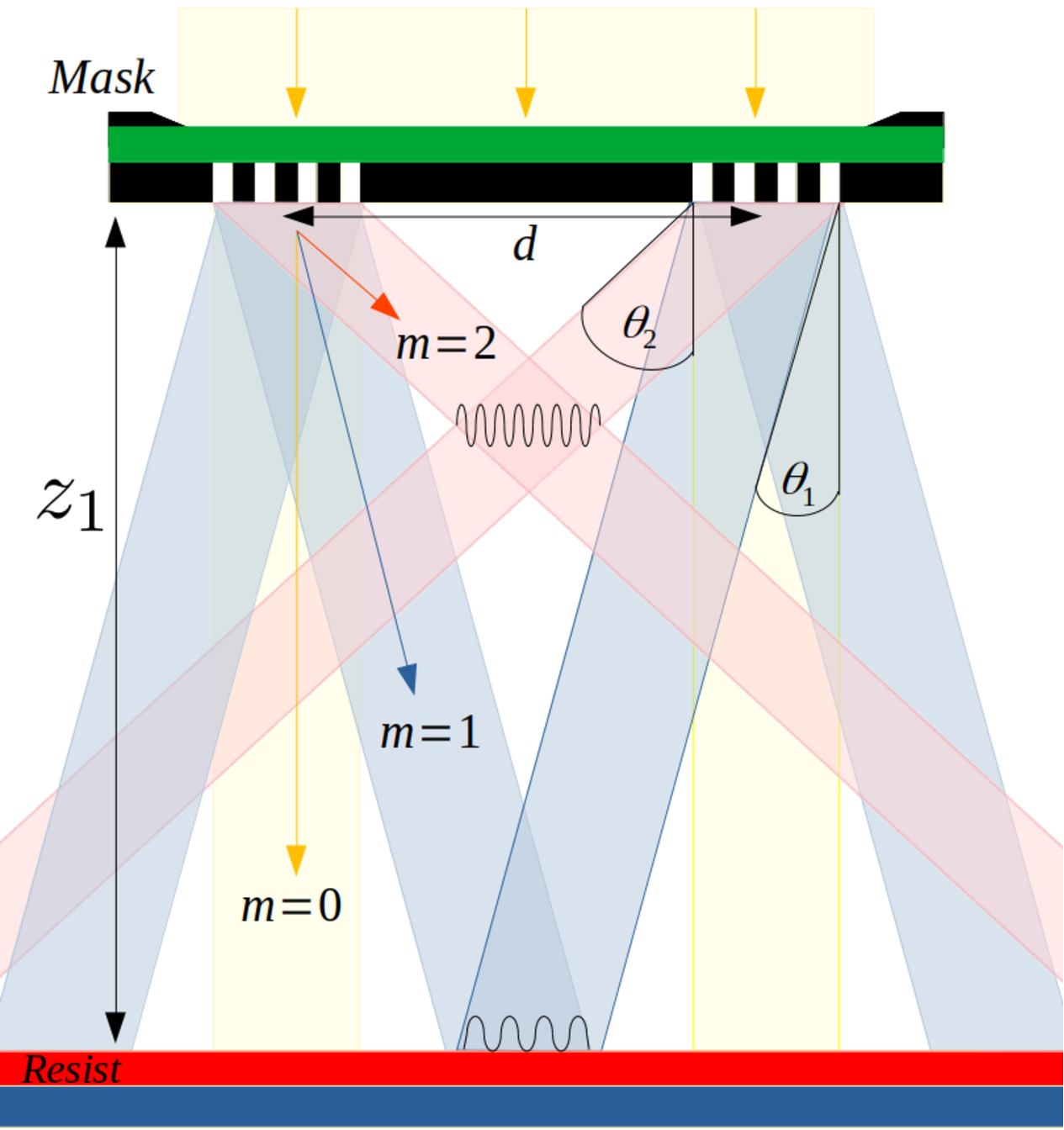}
    \caption{Schematic of an interference lithography setup using a binary grating mask. The first order ($m$= $\pm$1) beams form an aerial image at the image plane, a distance $z_1$ from the mask, which is then transferred onto a photoresist. The zero order ($m$=0) and second order ($m$=2) diffracted beams are also shown.}
    \label{figure:InterferenceLithography}
\end{figure}

As shown in \autoref{figure:InterferenceLithography}, the $m^{th}$ order beams from each grating will interfere at the aerial image plane at a distance $z_m$ from the mask, given by:
\begin{equation}
    z_m = \frac{d}{2\tan{\theta_m}} = \frac{d\sqrt{p_G^2 - m^2\lambda^2}}{2m\lambda}.
    \label{e:AerialImageDistance}
\end{equation}
For the simplest IL setup, consisting of two plane waves intersecting at a $2\theta_m$ angle, the aerial image will have a period $p$, given by
\begin{equation}
    p = \frac{\lambda}{2\sin{\theta_m}}.
    \label{e:AerialImagePeriod}
\end{equation}
The standard measure of the resolution of an aerial image is the half-pitch, HP=$p/2$. For a grating illuminated at normal incidence \autoref{e:GratingEquation} and \autoref{e:AerialImagePeriod} can be combined to show that
\begin{equation}
    \text{HP} = \frac{\lambda}{4\sin{\theta_m}} = \frac{p_G}{4m}.
    \label{e:AerialImageHalfPitch}
\end{equation}
The ultimate resolution for a first order aerial image is therefore $p_G/4$ \cite{karim_high-resolution_2015}. \autoref{e:AerialImageHalfPitch} implies that sub-10~nm patterning by IL requires a grating mask with $p_G < 40$~nm, which can be readily fabricated using electron beam lithography \cite{vieu_electron_2000}. It also shows the achromatic nature of IL, as the right side of \autoref{e:AerialImageHalfPitch} is independent of $\lambda$.

\subsection{Source Model}
\label{s:Source-Model}

Synchrotron Radiation Workshop (SRW) \cite{chubar_accurate_1998} was used to construct a model of branch B the SXR beamline at the AS, shown in \autoref{fig:beamline}~\cite{knappett_wavefront_2021}. A new model discussed in the present work includes the proposed EUV-IL endstation, permitting evaluation of the influence of all beamline parameters on the properties of the aerial image. SRW is commonly used to model SXR/EUV synchrotron beamlines and is capable of fully or partially coherent propagation \cite{he_wave_2020,meng_analysis_2021}.

\subsubsection{Storage Ring and Electron Beam:} 
\label{s:Storage-Ring-And-Electron-Beam}

The parameters used to model the electron beam and storage ring were obtained from Wootton \textit{et al.}~\citeyear{wootton_observation_2012} and Wootton \& Rassool~\citeyear{wootton_apple-ii_2013}, and are listed in \autoref{table:storageRing}. 

\begin{table}
\begin{center}
\caption{Key values for the Australian Synchrotron storage ring. Parameters obtained from \cite{wootton_observation_2012}}
\resizebox{0.5\textwidth}{!}
{\begin{tabular}{llr}
\hline
             & Parameter                & Value        \\ \hline
$E_0$        & Beam Energy        & 3.01 GeV         \\
$I_0$        & Current               & 0.2 A         \\
$\sigma_E$   & RMS Energy Spread       & 0.1021 $\%$   \\
$\epsilon_x$ & Horizontal Emittance   & 10 nm rad   \\
$\epsilon_y$ & Vertical Emittance  & 0.009 nm rad  \\
$\beta_x$    & Horizontal Beta Function & 9 m          \\
$\beta_y$    & Vertical Beta Function  & 3 m          \\ \hline
\end{tabular}}
\end{center}
\label{table:storageRing}
\end{table}

\subsubsection{APPLE II Undulator:}
\label{s:Apple-II-Undulator}

The SXR beamline utilises an elliptically polarising APPLE-II undulator source. This undulator was modelled as ideal, with magnetic fields defined by the undulator length, \textit{L}, magnetic period, $\lambda_u$, and number of periods, $N_u$. The undulator parameters were obtained from Wootton \textit{et al.}~\citeyear{wootton_observation_2012} and Wootton \& Rassool~\citeyear{wootton_apple-ii_2013}, and are shown in \autoref{table:undulator}. 

\begin{table}
\begin{center}
\caption{Key values for the APPLE-II undulator used for the SXR beamline at the Australian Synchrotron. $K_u$, which determines $B_u$, was adjusted to produce a fundamental harmonic at 184.76 eV. Other values obtained from Ref. \cite{wootton_observation_2012}}
\resizebox{0.5\textwidth}{!}{\begin{tabular}{llr}
\hline
            & Parameter                   & Value        \\ \hline
$L_u$       & Undulator Length           & 1.875 m     \\
$\lambda_u$ & Undulator Period Length     & 75 mm       \\
$N_u$       & Number of Undulator Periods & 25              \\
$B_u$       & Peak Magnetic Field         & 0.4611 T         \\
$K_u$       & Deflection Parameter        & 3.230           \\ \hline
\end{tabular}}
\end{center}
\label{table:undulator}
\end{table}

\subsection{Beamline Model}
\label{s:Beamline-Model}

\begin{figure}
    \centering
    \includegraphics[width=0.68\textwidth]{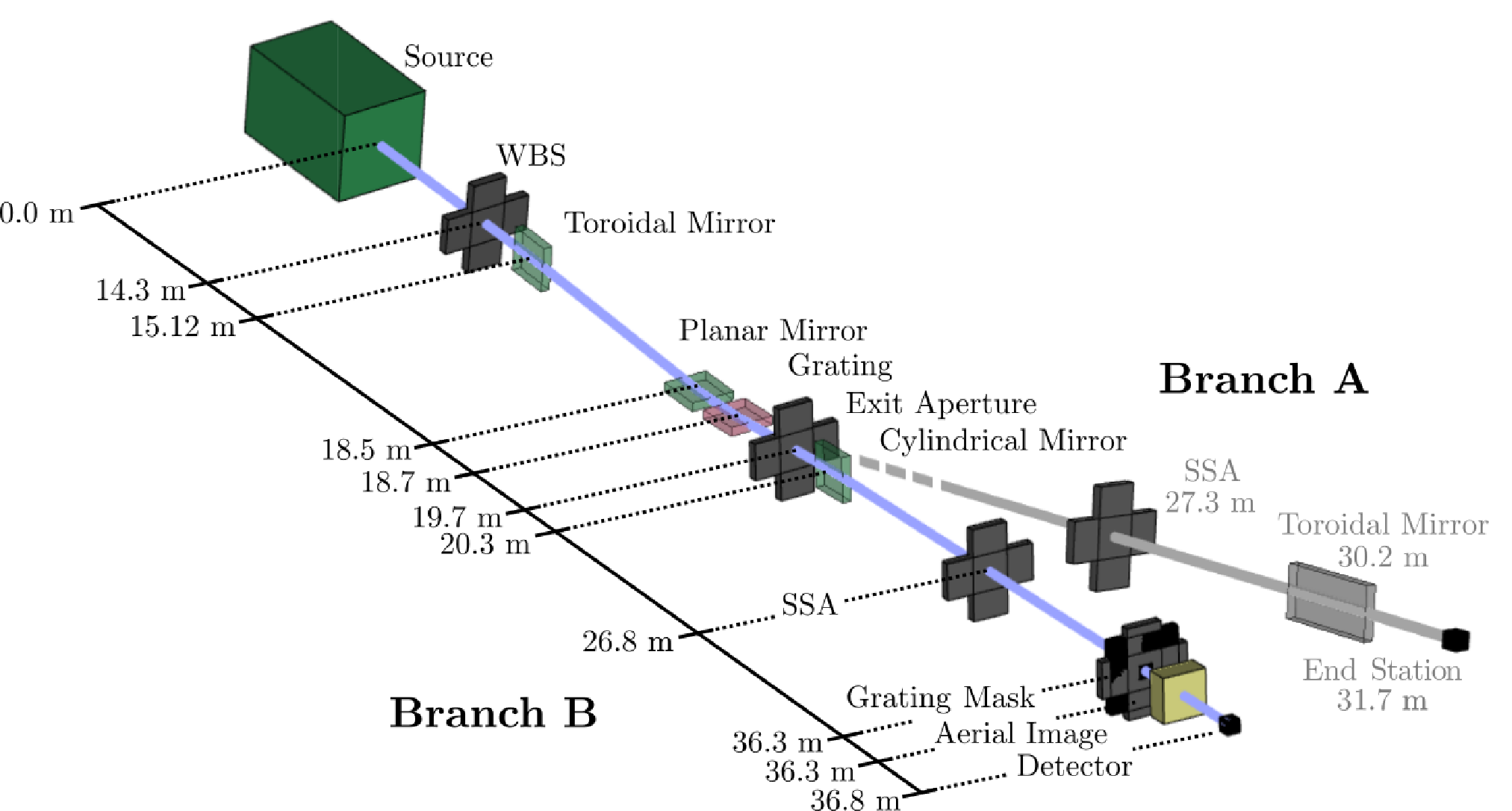}
    \caption{Schematic illustration of the SXR beamline. The distance of each element from the source is indicated (not to scale). Proposed IL optics and an imaging detector are included in branch B. Branch A was used to obtain measurements of the undulator harmonic intensity using photoelectron spectroscopy.}
    \label{fig:beamline}
\end{figure}

A schematic of the optical elements in use for each branch of the SXR beamline is shown in \autoref{fig:beamline}. The key optical components include white beam slits (WBS), a toroidal mirror, a planar grating monochromator (PGM) and a cylindrical mirror that can be rotated to direct the beam to branch A or B. The schematic shows the proposed location of IL optics in branch B, along with an existing EUV/SXR area detector. For typical operating conditions of branch B, the WBS are wide open (not touching the beam) and the bandwidth of the beam is defined by the PGM and secondary source aperture (SSA). The PGM disperses the monochromatic components of the incident beam at angles dependent on the photon energy as defined in \autoref{e:GratingEquation}, the beam is then focussed at the SSA, the size of which defines the resolving power of the beam as $(\lambda/ d \lambda)$.
All relevant parameters are listed in \autoref{table:beamline}. 

\begin{table}
\caption{The typical geometry of the optical elements used in the model of the beamline and IL optics. The defining characteristic for a mirror is given as the major radius, \textit{R}, and minor radius, $\rho$. The defining characteristic for a grating is the periodicity, and for a mask the grating size, $G$. Aperture sizes shown were used for all simulations and measurements shown in this work unless stated otherwise.}
\resizebox{\textwidth}{!}{\begin{tabular}{lrrrrrr}
\hline
\textbf{Element Name}     & \textbf{\begin{tabular}[c]{@{}r@{}}Element\\ Type\end{tabular}} & \textbf{\begin{tabular}[c]{@{}r@{}}Propagation\\ Distance [m]\end{tabular}} & \textbf{\begin{tabular}[c]{@{}r@{}}Distance from\\ Source [m]\end{tabular}} & \textbf{Dimensions [mm]}   & \textbf{\begin{tabular}[c]{@{}r@{}}Defining\\ Characteristics\end{tabular}} & \textbf{\begin{tabular}[c]{@{}r@{}}Incident\\ Angle [$^o$]\end{tabular}} \\ \hline
White Beam Slits          & Aperture                                                        & 14.30                                                                 & 14.30                                                                 & 4 $\times$ 3       & N/A                                                                         & 90                                                            \\
Toroidal Mirror           & Mirror                                                          & 0.82                                                                 & 15.12                                                                 & 420 $\times$ 30    & \begin{tabular}[c]{@{}r@{}}R = 6668 m\\ $\rho$ = 5.262 m\end{tabular}    & 1                                                             \\
Planar Mirror             & Mirror                                                          & 3.38                                                                  & 18.50                                                                 & 460 $\times$ 50    & N/A                                                                         & 2                                                             \\
Grating                   & Grating                                                         & 0.20                                                                  & 18.70                                                                 & 150 $\times$ 20    & 250 lines/mm                                                                & 1.173                                                        \\
Exit Aperture             & Aperture                                                        & 1.00                                                                  & 19.70                                                                 & 10 $\times$ 20     & N/A                                                                         & 90                                                            \\
Cylindrical Mirror        & Mirror                                                          & 0.60                                                                  & 20.30                                                                 & 240 $\times$ 40    & \begin{tabular}[c]{@{}r@{}}R = 100 km\\ $\rho$ = 0.2443 m\end{tabular}      & 1.5                                                           \\
Secondary Source Aperture & Aperture                                                        & 6.50                                                                  & 26.80                                                                 & 0.025 $\times$ 0.025   & N/A                                                                         & 90                                                            \\
Grating Mask              & Mask                                                            & 9.50                                                                  & 36.30                                                                 & 0.15 $\times$ 0.15 & \begin{tabular}[c]{@{}r@{}}$G$ = 50 \textmu m\end{tabular}  & 90                                                            \\ \hline
\end{tabular}}
\label{table:beamline}
\end{table}

\subsection{Partially Coherent Simulations}
\label{s:Partially-Coherent-Simulations}

Synchrotron Radiation Workshop (SRW) implements fully spatially coherent wavefront propagation through the propagation of monochromatic radiation emitted by a single relativistic electron passing through an undulator. The wavefield is represented by the monochromatic, transverse components of the electric field $E_{\perp \lambda}$ and paraxial propagation is done through discrete FFTs \cite{chubar_accurate_1998}. Taking $x$ and $y$ as the horizontal and vertical directions respectively, and $z$ as the direction along the beam axis, the propagated wavefield, $E_{\perp \lambda}(x,y,z)$ from each wavefield component at $z$=0, $E_{\perp \lambda}(x,y,z=0)$ can be written as
\begin{equation}
    \begin{split} 
    E_{\perp \lambda} (x,y,z)
    & = -\frac{ik \exp (ikz)}{2 \pi z} \int\int_{-\infty}^{\infty} E_{\perp \lambda} (x',y',z=0) \\
    & \times \exp \left [ \frac{ik}{2z} [(x-x')^2 + (y-y')^2)] \right ] dx'dy' ,
    \end{split}
    \label{e:WavefrontPropagation}
\end{equation}
where $k=\sqrt{k_x^2 + k_y^2 + k_z^2}$ is the spatial frequency of the plane wave component over an area of phase space $dk_x dk_y$. In all simulations shown in this work, only a single monochromatic component is considered, where the spatial frequency $k=\frac{2\pi}{\lambda}$, and a quadratic approximation to the exponential phase term in \autoref{e:WavefrontPropagation} \cite{chubar_memory_2019} was used where appropriate. 

Partially coherent, monochromatic propagation is achieved by randomly sampling the phase space of the entire electron beam using the Monte-Carlo method \cite{laundy_partial_2014}. The number of electrons needed to accurately represent the electron beam increases with the electron beam emittance \cite{chubar_initial_2016}. For all simulations shown in this work, 4000 electrons were sufficient to accurately model the monochromatic beam \cite{knappett_wavefront_2021}. The electric field of each electron is then propagated coherently using \autoref{e:WavefrontPropagation} and the partially coherent electric field is calculated as the average of the electric field distributions from each electron, seeded over the phase space occupied by the beam \cite{chubar_recent_2014}.

\section{Simulation Results}
\label{s:Simulation-Results}

\subsection{Beam Profile}
\label{s:Sim:Beam-Profile}

The beamline model was used to propagate a partially coherent wavefield from the source through to the grating mask plane at $\lambda =$ 4.96, 6.7, 9.2, and 13.5~nm. The intensity was modelled from the complex electric field as $I(x,y)=|E(x,y)|^2$, with the pixel size chosen such that it satisfies the Nyquist-Shannon sampling theorem to ensure satisfactory representation of phase \cite{shannon_communication_1949}. 
The full-width at half-maximum (FWHM) of the intensity profile at the mask plane was calculated for each wavelength. 

\subsection{Flux and Coherence at the Mask Plane}
\label{s:Flux-And-Coherence-At-Mask-Plane}

\begin{figure}
    \centering
    \includegraphics[width=0.5\textwidth]{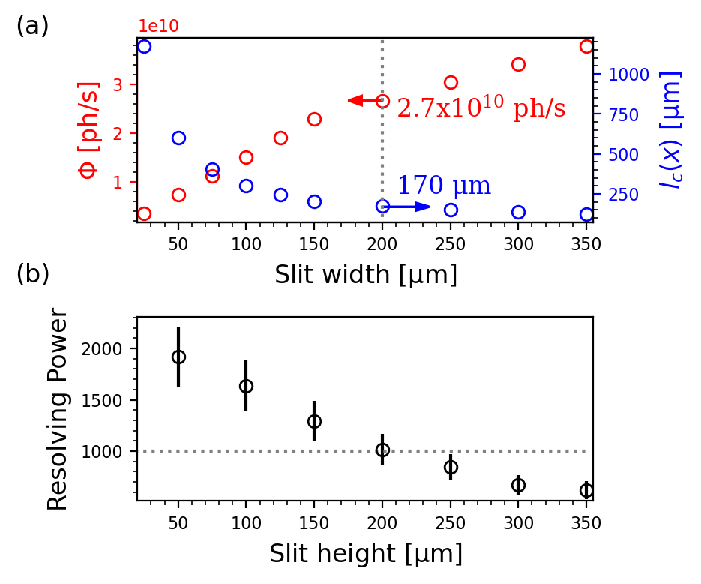}
    \caption{(a) The total photon flux (red) horizontal coherence length (blue), and (b) the resolving power at the mask plane of beamline branch B for different SSA sizes. 
    The resolving power is shown as a function of SSA height because the energy resolution of the PGM is defined only by the vertical SSA size. The grey dotted lines correspond to the maximum SSA size that provides coherent illumination of a representative grating mask (see Figure \ref{fig:FluxAndIntensity}).}
    \label{fig:FluxvsCoherence}
\end{figure}

The total flux and spatial coherence of the wavefield at the mask plane was evaluated after partially coherent propagation through the beamline model. The total flux, $\Phi$ was calculated in units of photons per second per 0.1\% bandwidth from the propagated complex electric field $E (x,y)$ by,
\begin{equation}
    \Phi \left [\text{ph/s/0.1\% bw} \right ]  = dxdy \times 10^{-4}  \sum_x \sum_y |E (x,y)|^2,
\end{equation}
where $dx$ and $dy$ are the pixel size of $E$ in $x$ and $y$ respectively. As SRW does not take into account mirror reflectivity or grating efficiency, the final flux values were adjusted to account for the reflectivity of each mirror and grating used in the PGM, shown in \autoref{table:beamline}. Since the reflection grating efficiency at EUV/SXR wavelengths is not accurately known, a constant efficiency of 10\% was assumed for all simulations, with a third order efficiency of 0.1\%.

The spatial coherence length, $l_c$ at the mask plane was calculated by first computing the one-dimensional mutual intensity $J$ using
\begin{equation}
    J(x_1,x_2) = \langle E^* (x_1) E(x_2)\rangle,
    \label{e:MutualIntensity}
\end{equation}
where $x_1$, and $x_2$ are different sets of points in the central horizontal axis $(x,y$=$0)$, so that when $x_1$=$x_2$, $J$ reduces to the horizontal intensity profile. The degree of coherence between any point along $(x,y=0)$ and the central point $(x$=$0,y$=$0)$ was then calculated following Meng \textit{et al.}~\citeyear{meng_analysis_2021}, using
\begin{equation}
    \gamma = \frac{J(x,0)}{[\langle|E(x)|^2\rangle \langle|E(0)|^2\rangle]^{1/2}}.
    \label{e:DegreeOfCoherence}
\end{equation}
The coherence length $l_c (x,y)$ can then be determined as the distance from the central point at which $\gamma<0.8$. Repeated propagations were undertaken for different horizontal SSA sizes, ranging from 25 \textmu m to 350 \textmu m. The results of the simulations are shown in \autoref{fig:FluxvsCoherence}, which shows that the total flux at the mask plane increases linearly with horizontal SSA size from 2.9$\times 10^{10}$ ph/s to 2.7$\times 10^{9}$ ph/s, while the coherence length at the mask plane decreases exponentially, from $\sim1.17$~mm to $\sim125$~\textmu m. Previous unpublished measurements by the authors using a Young's double-slit array with a maximum slit separation of 30 \textmu m
showed no significant loss of visibility with any horizontal SSA size, indicating that the horizontal coherence length at 6.7~nm wavelength is always greater than 30 \textmu m.

The bandwidth was measured using X-ray photoelectron spectroscopy (XPS) of the Au Fermi edge. The Fermi edge broadening was measured with different vertical SSA sizes and the energy bandwidth of the fundamental undulator harmonic at $\lambda$=6.7~nm was estimated for each SSA size. The results are shown in \autoref{fig:FluxvsCoherence}.

\section{Experimental Measurements}
\label{s:Experimental-Measurements}

\begin{figure}
    \centering
    \includegraphics[width=0.5\textwidth]{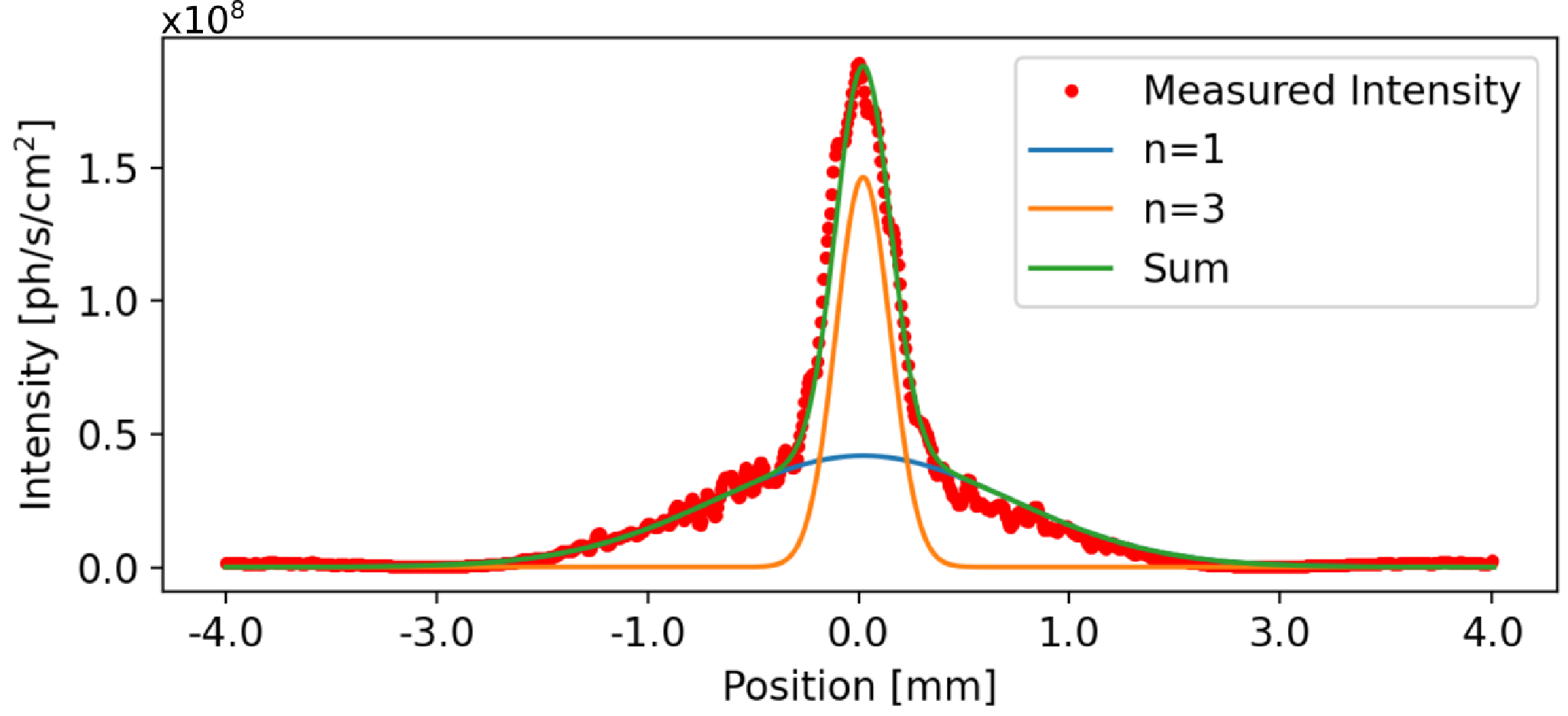}
    \caption{The measured horizontal intensity profile, 50 cm from the BDA with a fundamental photon energy of 185~eV and $C_{ff}=2$. The horizontal line profile through the beam, overlaid with a fitting of two Gaussians representing the fundamental (harmonic $n$=1) and the third harmonic ($n$=3).}
    \label{fig:IntensityProfileWithFitting}
\end{figure} 

Measurements of the intensity profile of the beam were taken at branch B of the beamline (\autoref{fig:beamline}), using an AXIS-SXR sCMOS detector with an EUV-Enhanced GSENSE400 BSI sensor \cite{harada_high_2019}. The detector was situated $z=0.5$~m from the beam defining aperture (BDA) plane. At the BDA plane, a 4.064 \textmu m thick Ultralene (C$_3$H$_6$) filter was positioned in the beam path to attenuate the beam and avoid over-saturating the detector. For each measurement, the WBS were set to $0.84\times1.0$~mm$^2$ $(x,y)$, and the SSA was set to $25\times25$~\textmu m$^2$ $(x,y)$. The PGM fixed focus constant ($C_{ff}$) was set to 1.4 instead of the standard operating value of 2, as the higher harmonic content ($\zeta$) is reduced for small $C_{ff}$ values \cite{kleemann_design_1997}, all other parameters were as shown in \autoref{table:beamline}. These small slit sizes mean all results presented in this section show less flux and intensity than can be expected with beamline parameters optimised for high flux. The photon energy was varied from 90~eV to 270~eV in 10~eV steps, with extra steps at 92~eV (13.5~nm) and 185~eV (6.7~nm). 
One hundred exposures were taken at each photon energy which were processed in sets of 10. Horizontal and vertical profiles across the summed intensity in each set were fitted with two Gaussians, as shown in Figure \ref{fig:IntensityProfileWithFitting}, that represent the fundamental (harmonic $n=1$) intensity profile, and the intensity profile of the higher ($n>3$) harmonics present in the beam. For the fitted Gaussian representing the fundamental, the FWHM was calculated (\autoref{fig:FWHM}). The total photon flux of each harmonic $\Phi_n$, was also calculated, as well as the intensity over a typical IL grating mask area, $I^G_n$. The grating mask area was chosen as a 4-grating mask, with each grating occupying a $50\times50$~\textmu m$^2$ area, arranged radially around the beam center, with a $50\times50$~\textmu m$^2$ central beam stop. A representation of the grating mask area over the two-dimensional beam intensity is shown in Figure \ref{fig:FluxAndIntensity}. Errors for each fitted parameter have been taken as the standard deviation of the fitted value for each set of 10 processed images. For each experimentally measured beam parameter discussed henceforth, equivalent parameters extracted from the partially coherent simulated intensity at 90.44, 135, 184.76, and 250~eV have been included for comparison.

The total flux was measured again as a function of photon energy using the AXUV100 photodiode at branch A (\autoref{fig:beamline}). Photodiode measurements were adjusted to account for the reflectivity of the extra mirror used in branch A. Photodiode measurements were taken with an SSA size of $14000$~\textmu m horizontal (\textit{i.e.}, wide open) $\times~20$~\textmu m vertical (branch A), while intensity measurements were taken with SSA $25 \times 25$~\textmu m$^2$ (branch B). The difference in horizontal SSA size between branch A and branch B lead to a 100 fold increase in flux at the photodiode, which has been accounted for by scaling the photodiode flux values in \autoref{fig:FluxAndIntensity} accordingly.

Measurements of the harmonic content of the beam were taken using XPS of Au 4f$^{1/2}$ peaks corresponding to different undulator harmonics. Through analysis of the relative peak areas, an estimate of $\zeta$ was calculated for fundamental photon energies from 130~eV to 350~eV in 20~eV steps:
\begin{equation}
    \zeta = \frac{\Phi_{n>1}}{\Phi_n},
\end{equation}
where $\Phi_{n>1}$ is the photon flux contained in higher undulator harmonics and $\sum_n \Phi_n$ is the total flux.

\section{Experimental Results}
\label{s:Experimental-Results}

\subsection{Beam Profile}
\label{s:Exp:Beam-Profile}

\begin{figure}
    \centering
    \includegraphics[width=0.5\textwidth]{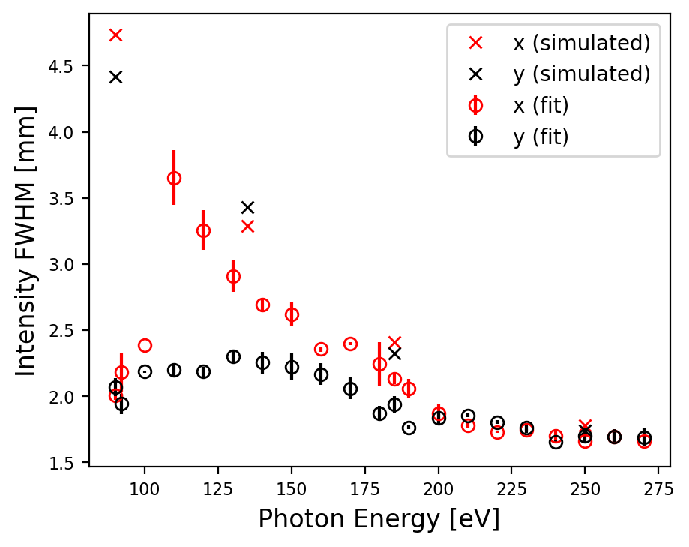}
    \caption{The intensity FWHM measured in the horizontal (red) and vertical (black) direction for photon energies from 90~eV to 270~eV and $C_{ff}$=1.4. 
    The calculated FWHM of intensity profiles generated through partially coherent wavefront propagation through the beamline model are also shown for photon energies of 90.44~eV, 135~eV, 184.76~eV, and 250~eV.}
    \label{fig:FWHM}
\end{figure}

The FWHM of the fundamental ($n$=1) intensity obtained from the fit to the measured intensity distribution at the grating mask plane is shown in \autoref{fig:FWHM} for photon energies 90~eV to 270~eV. Comparison with simulated intensity FWHM shows a similar dependence on photon energy. However at low photon energies the comparison between the two methods shows less agreement, which may be attributed to uncertainties resulting primarily from the low signal-to-noise for the intensity measurements taken with low photon energy.

\subsection{Higher Harmonic Content}
\label{s:Higher-Harmonic-Content}

\begin{figure}
    \centering
    \includegraphics[width=0.5\textwidth]{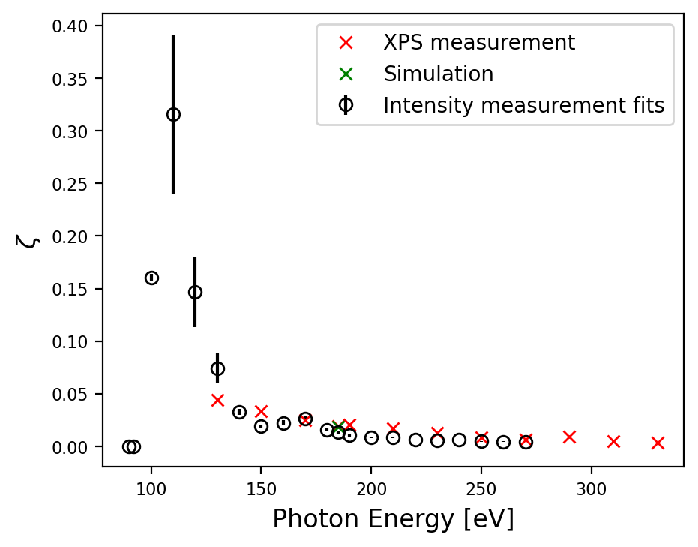}
    \caption{Higher harmonic content ($\zeta$) measurements using direct beam measurements (black) and XPS measurements (red). A single calculation from simulation has been included at 184.76~eV (green).}
    \label{fig:HarmonicContaminationPlots}
\end{figure}

Measurements were taken by XPS with $C_{ff}$ values of 2 and 1.4. The total $\zeta$ was found to significantly reduce at $C_{ff}=1.4$ for all photon energies measured as expected \cite{kleemann_design_1997}, with a reduction from $\sim10$\% harmonic content at $C_{ff}=2$ to $\sim2$\% at $C_{ff}=1.4$ for a fundamental photon energy of 185~eV. The results of the harmonic content measurements are shown in \autoref{fig:HarmonicContaminationPlots} for $C_{ff}=1.4$ and compared to the harmonic contamination given by the fitting of Gaussians to the direct intensity measurements. The XPS measurements show good agreement with the direct beam measurements when comparing total intensity, and the simulation results show close agreement to both measurements. 

\subsection{Flux and Intensity at the Mask Plane}
\label{s:Flux-And-Intensity-At-Mask-Plane}

\begin{figure}
    \centering
    \includegraphics[width=0.5\textwidth]{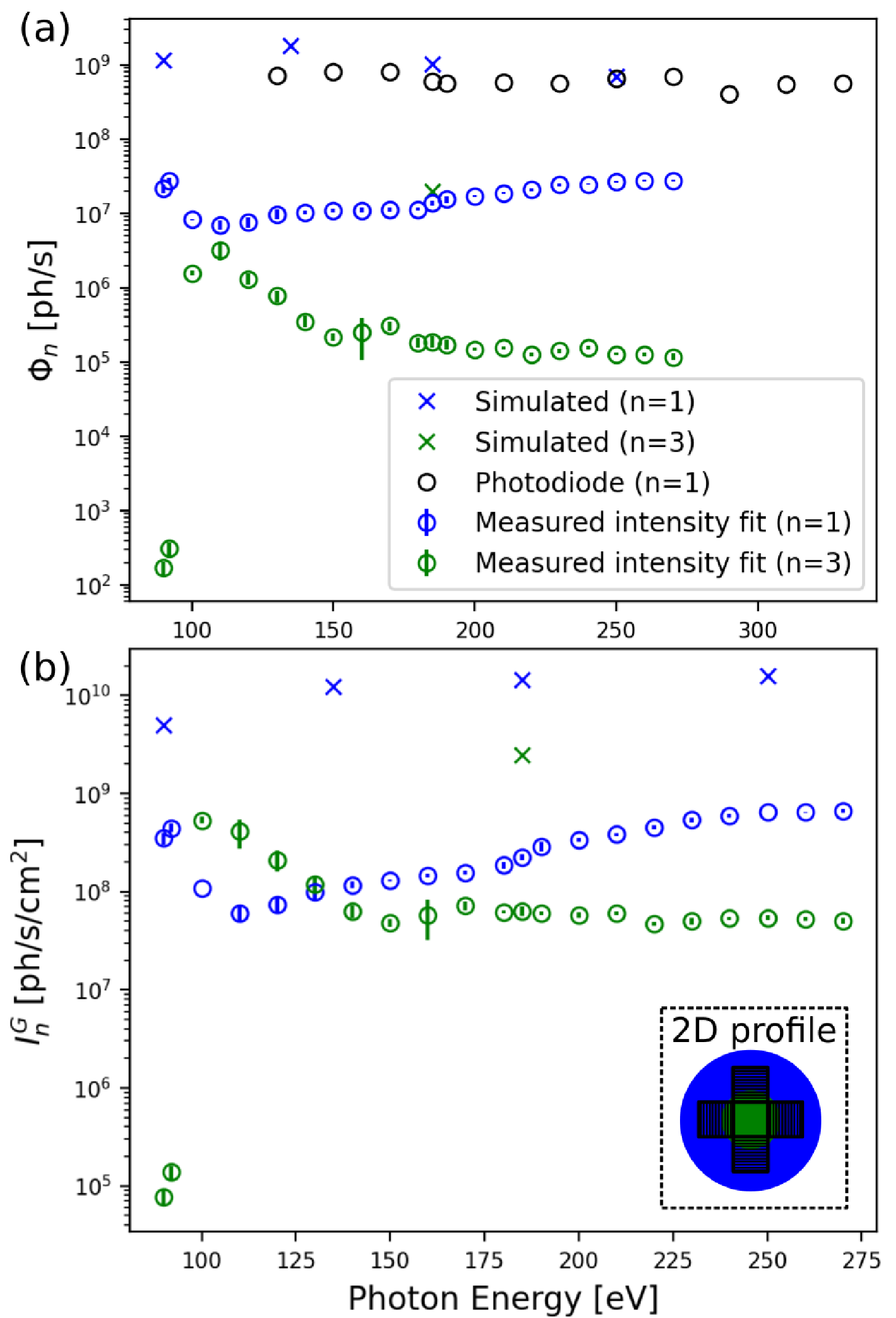}
    \caption{(a) The simulated and measured total flux ($\Phi_n$) and (b) the total intensity over the area of a 4-grating mask ($I_n^G$) consisting of $50\times50$~\textmu m gratings arranged symmetrically with a 50 \textmu m gap in the center (shown in inset). Flux and intensity measurements were taken at branch B, while flux measurements were also taken on branch A using an AXUV100 photodiode. $C_{ff}$=1.4 for all measurements shown. 
    }
    \label{fig:FluxAndIntensity}
\end{figure}

The total flux as a function of photon energy is shown in \autoref{fig:FluxAndIntensity} as measured by three different methods; the AXUV100 photodiode at branch A beamline (\autoref{fig:beamline}), the fundamental fitting of the direct beam intensity measurements taken at branch B and simulated wavefront propagation.
The intensity measured over a typical IL-grating area, $I^G_n$, described earlier and shown in Figure \ref{fig:FluxAndIntensity}, is also included for each photon energy.
The simulated results show significantly greater flux and intensity over the grating area compared to the direct intensity measurements, however they show close agreement to the flux measurements using the photodiode.
\autoref{fig:FluxAndIntensity} also shows that the intensity-on-mask of $n$=1 at 185~eV is $I^G_1\sim2\times10^{8}$~ph/s/cm$^2$, which gives a dose-on-mask of just $D_m\sim6\times10^{-6}$~mJ/s/cm\textsuperscript{2}. However, recall that the intensity measurements shown in \autoref{fig:FluxAndIntensity} were made with `low flux' beamline settings, using small WBS and SSA sizes to avoid over-saturation of the detector. Using `high flux' settings, the intensity at the mask plane is expected to be closer to the simulated results shown in \autoref{fig:FluxvsCoherence}. Hence, the results shown in \autoref{fig:FluxAndIntensity} serve only to confirm that the simulation can accurately predict intensity at the mask plane.

To evaluate the beamline's suitability for IL, partially coherent wavefront propagation was used to calculate $\Phi_1$ and $I^G_1$ as a function of SSA size for photon energy 185 eV.  The SSA height and width were kept equal and varied from 200 \textmu m to 1500 \textmu m in 100 \textmu m steps. $I^G_1$ was found to reach a maximum of 5.85$\times 10^{13}$ ph/s/cm$^2$ for a 300$\times$300 \textmu m SSA.   For the same representative grating with efficiency $\eta=6\%$  illustrated in Figure \ref{fig:FluxAndIntensity}, this value of $I^G_1$ corresponds to a value of $D_w$ equal to 5.46 mJ/s/cm$^2$.  For $D_m =11$~mJ/cm\textsuperscript{2}, the minimum exposure time is then 1.9~s. The coherence length at the mask plane for these settings is 137 \textmu m, which would allow for the coherent illumination of a mask containing 45 \textmu m gratings.

\section{Conclusion}
\label{s:Conclusion}

Partially coherent wavefront propagation simulations have been used to accurately model the SXR beamline  at the Australian Synchrotron. The model was shown to give a quantitative representation of photon flux that agrees with direct experimental measurements using a AXUV100 photodiode. However, the simulated flux and the photodiode flux measurements did not show agreement with the direct intensity measurements. This could be due to an error in the thickness and density of the Ultralene filter used to attenuate the beam, which has been found to vary in thickness by 10\% from the nominal value of 4~\textmu m \cite{surowka_model_2020}. The discrepancy could also have come from errors in simulation due to the assumption of ideal beamline elements, reflectivities and diffraction efficiencies, or a possible misalignment of the SSA in branch B of the SXR beamline, causing a drop in flux at the detector plane.

A method for higher harmonic suppression by adjusting the $C_{ff}$ of the PGM from 2 to 1.4 was found to reduce total harmonic contamination at the mask plane from $\sim10$\% to $\sim2$\%.

The beamline was shown to possess adequate power and coherence for IL at EUV/SXR wavelengths when operating with a WBS of $4 \times 3$~mm$^2$ and an SSA of $300 \times 300$~\textmu m$^2$, allowing for 1.9~s patterning of a 45 \textmu m maximum grating size, which is compatible to that used at other EUV-IL facilities \cite{sahoo_development_2023}. The exposure time is sufficiently short for maintaining the nanometre mechanical stability required for studying lithographic processes at high spatial resolution. Greater power is possible at the cost of coherence---and therefore grating size---with experimental evidence that the transverse coherence length exceeds 30 \textmu m for any choice of SSA size. 

\ack{Acknowledgements}

This research was supported by the Australian Research Council (ARC) Linkage Infrastructure, Equipment and Facilities (LIEF) scheme, which funded part of the equipment used in this work. One of the authors (JBMK) was supported by an Australian Government Research Training Program (RTP) Scholarship, and an AINSE Ltd. Postgraduate Research Award (PGRA). Synchrotron experiments were carried out at the SXR beamline of the Australian Synchrotron, part of ANSTO. The authors would like to thank Trey Guest for his early contributions to beamline modeling and helpful discussions.

\referencelist[ref]

\end{document}